\documentclass[preprint,showpacs,preprintnumbers,amsmath,amssymb,nofootinbib]{revtex4}
\usepackage{graphicx}
\usepackage{dcolumn}
\usepackage{bm}
\usepackage{amssymb}
\usepackage{epsfig}

\newcommand\figwidth{3in}
\newcommand\plotwidth{.6\textwidth}

\newcommand\half{\frac{1}{2}} 
\newcommand\beq{\begin{eqnarray}}
\newcommand\eeq{\end{eqnarray}}

\newcommand\Eq[1]{Eq. (\ref{eq:#1})}
\newcommand\Sec[1]{Sec. (\ref{sec:#1})}
\newcommand\Fig[1]{Fig. (\ref{fig:#1})}

\newcommand{\bfn}{{\bf n}}
\newcommand{\bfm}{{\bf m}}
\newcommand{\bfe}{{\bf e}}
\newcommand{\calB}{{\cal B}}
\newcommand{\calC}{{\cal C}}
\newcommand{\calI}{{\cal I}}
\newcommand{\calL}{{\cal L}}
\newcommand{\calO}{{\cal O}}
\newcommand{\calN}{{\cal N}}
\newcommand{\mathZ}{{\mathbb Z}}
\newcommand{\mathN}{{\mathbb N}}

\newcommand{\Nt}{N_\tau}
\newcommand{\Ns}{N_s}
\newcommand{\la}{\langle}
\newcommand{\ra}{\rangle}

\begin{document}

\preprint{INT-PUB 06-16}

\title{Method for simulating $O(N)$ lattice models at finite density}
\author{Michael G. Endres}
\email{endres@u.washington.edu}
\affiliation{Institute for Nuclear Theory, University of Washington, Seattle, WA 98195-1550}

\begin{abstract}
We present a method for simulating relativistic and nonrelativistic scalar field theories at finite density, with matter transforming in the fundamental representation of the global symmetry group $O(N)$.
The method avoids the problem of complex probability weights which is present in conventional path integral Monte Carlo algorithms.
To verify our approach, we simulate the free and interacting relativistic $U(1) \simeq O(2)$  theory in $2+1$ dimensions.
We compute the two-point correlation function and charge density as a function of chemical potential in the free theory.
At weak $|\phi|^4$ coupling and zero temperature we map the $m^2-\mu$ phase diagram and compare our numerical results with perturbative calculations.
Finally, we compute properties of the $T-\mu$ phase diagram in the vicinity of the phase transition and at bare self-couplings large compared to the temperature and chemical potential.
\end{abstract}
\date{\today}

\pacs{} 

\maketitle

\section{Introduction}
\label{sec:0}
Monte Carlo simulations provide a reliable method for studying nonperturbative physics of quantum field theories.
It is unfortunate that with few exceptions, path integral Monte Carlo techniques are not suitable for the study of relativistic systems at finite density.
In the path integral formalism, standard importance sampling techniques rely on the interpretation of the integrand $e^{-S[\Phi]}$ as a probability weight associated with the field configuration $\Phi$.
At nonzero chemical potential, the bosonic and fermionic Euclidean actions are typically complex, rendering importance sampling inapplicable.
This problem is commonly known  as the ``sign problem''.
\footnote{Some specific examples of models that avoid the sign problem include QCD at finite isospin density \cite{Alford:1998sd}, two-color QCD \cite{Kogut:2001na}, and low energy fermions with attractive interactions \cite{Chen:2003vy}.}
In many instances, the sign problem persists in the path integral formulation of nonrelativistic theories as well--both at finite and at zero chemical potential.
In these cases, this sign problem does not arise because the system is at finite density, but rather because the action is linear in time derivatives.

Numerous schemes have been proposed for numerical study of relativistic theories at finite density, with emphasis on lattice QCD.
Reweighting \cite{Ferrenberg:1988yz} has had limited success for models in the regime $\beta \mu < 1$.
Here, one folds the phase of the probability measure into expectation values of observables.
Statistical ensembles are then generated using the modulus of the complex measure.
For larger $\beta \mu$, the method fails because the calculation of expectation values involve intricate cancellation among phases, leading to overwhelming statistical errors.
In addition, one must contend with the possibility that ensembles generated in simulations have poor overlap with the true probability distribution.
In some cases, one may reduce the severity of the overlap problem with the use of multiparameter reweighting techniques \cite{Fodor:2001au}. 

Some have investigated theories at imaginary chemical potential, where lattice simulations are viable \cite{Alford:1998sd}. 
Using this approach, one may compute the canonical partition function for a system with total fixed charge $Q$ via the Fourier transform
\beq
\label{eq:FTZ}
Z_Q = \int_0^{2 \pi}\! d\theta \, Z(i \theta/\beta) e^{i Q \theta} \ .
\eeq
In practice, this approach only postpones the sign problem since numerical integration of \Eq{FTZ} relies on the cancellation of phases, which become increasingly severe at large values of $Q$.
A second option is to analytically continue the partition function to real $\mu$.
The analytic continuation is typically based upon a Taylor series expansion about $\mu=0$ and therefore this approach is only valid within the radius of convergence--inevitably failing at and beyond any critical point.
In addition, the analytic continuation is limited by the periodicity of the partition function $Z(i \mu) = Z(i \mu + i 2 \pi/\beta)$ along the imaginary $\mu$ axis--a property which is evident from analysis of the fugacity expansion.

In the case of scalar theories, which is the focus of this paper, we investigate the prospects of finding alternative representations for the partition function which are suitable for Monte Carlo simulations.
To this end, we formulate relativistic and nonrelativistic $O(N)$ scalar theories in terms of dual lattice variables \cite{Ukawa:1979yv}.
The method makes use of $U(1)$ character expansions to explicitly integrate out the angular mode of a complex scalar field.
In the dual representation, the angular mode is replaced by a set of integer valued link variables which may be interpreted as the conserved current density of the system.
The path integral formulation is particularly suited for Monte Carlo simulation because complex phases associated with the chemical potential (in the case of relativistic theories) and lattice time derivative (in the case of nonrelativistic theories) are avoided.
All physical observables are measurable within the formalism--free of sign problems--provided the corresponding operators are neutral in charge.
We will consider scalar theories in the grand canonical ensemble, although as we shall see, the formulation is applicable to canonical ensembles as well.

To motivate the approach, we begin by considering a quantum mechanical rigid rotor at finite temperature ($T=\beta^{-1}$) and chemical potential $\mu$ coupled to angular momentum.
This model corresponds to a $(0+1)$-dimensional relativistic scalar theory at finite charge density, where the dynamics of the heavy radial mode have been ignored (or integrated out).
The continuum partition function is defined by the path integral representation
\beq
Z(\mu) = \int\! [d\Sigma]\, e^{-\int_0^\beta\! d\tau\, \frac{\calI}{2} \left[\partial_\tau \Sigma^\ast \partial_\tau \Sigma+\mu (\Sigma^\ast \partial_t \Sigma-\Sigma \partial_t \Sigma^\ast)-\mu^2 \right]}\ ,
\eeq
where $\Sigma(\tau) = e^{i \theta(\tau)}$ is an element of $U(1)$ and $\calI$ is the moment of inertia.
At finite temperature, we impose periodic boundary conditions on $\Sigma$ in the imaginary-time direction.
The periodicity of $\Sigma$ in the imaginary-time direction translates into the boundary condition $\theta(\tau) = \theta(\tau+\beta) + 2\pi w$ on the angular mode, where $w$ is the winding number associated with the field configuration $\theta(\tau)$.
Our definition of the partition function implicitly includes a sum over all values of the winding number; this sum may be made explicit by introducing a new field $\theta_w(\tau)$, which represents configurations with fixed winding number $w$.
The partition function may then be written as:
\beq
Z(\mu) = \sum_{w\in\mathZ} \int\![d\theta_w]\, e^{-\int_0^\beta\! d\tau\, \frac{\calI}{2} \left[(\partial_\tau \theta_w)^2+ 2 i \mu \partial_t \theta_w -\mu^2 \right]}\ .
\eeq
By taking advantage of the correspondence between configurations with zero and nonzero winding number, namely $\theta_w(\tau) = \theta_0(\tau) + 2 \pi w \tau/\beta$, we may express the partition function as the product of two terms:
\beq
\label{eq:RigidRotor}
Z(\mu) =  \int\! [d\theta_0]\, e^{-\int_0^\beta\! d\tau\, \frac{\calI}{2} (\partial \theta_0)^2} \sum_{w\in\mathZ} e^{-\beta \frac{\calI}{2} \left[(2\pi w/\beta)^2+2i\mu (2\pi w/\beta)-\mu^2\right]}\ ,
\eeq
where the  chemical potential decouples from $\theta_0$ but remains coupled to winding number.
Since we are only interested in the $\mu$-dependence of the partition function, we may neglect the $\beta$-dependent path integral over the zero winding number sector of the theory.
Using the Poisson summation formula, the partition function is expressed as a sum over positive weights:
\beq
\label{eq:DualRigidRotor}
Z(\mu) \propto \sum_{Q\in\mathZ} e^{-\beta \left[\frac{Q^2}{2 \calI}+\mu Q\right]}\ .
\eeq
The resulting representation for the partition function is given in terms of the momentum conjugate variable to the coordinate $\theta$, namely the angular momentum Q, indicating that the sign problem is a basis dependent problem; this observation forms the basis for formulating the path integral for scalar field theories in terms of dual lattice variables.
In passing from \Eq{RigidRotor} to \Eq{DualRigidRotor}, the dimensionless ratio $\calI/\beta$ is mapped to $\beta/\calI$--a typical property of duality transformations--and the chemical potential now couples to total angular momentum (or charge) $Q$ of the system.

In \Sec{1} we will develop a generalized formalism for relativistic and nonrelativistic $O(N)$ scalar field theories defined on a space-time lattice and describe how to compute expectation values within this framework.
In addition, we outline standard procedures for solving certain continuity constraints that arise in our derivation.
For completeness, we show how $U(1)$ gauge fields fit into the formalism as well.
In \Sec{2} we explain several numerical algorithms for simulating relativistic bosonic theories at finite density and discuss additional issues that require resolution in order to render such simulations practical.
\Sec{3} pertains to the details of our simulations.
A comparison of numerical and analytic results is presented for the free and interacting $|\phi|^4$ theory in $2+1$ dimensions and at weak coupling.
\Sec{4} is devoted to an application of our methods to the same model at finite temperature and density, and in a nonperturbative regime.
Here, we study the behavior of the theory in the vicinity of the zero temperature and density critical line, and determine several universal constants associated with the phase transition.

\section{Formalism}
\label{sec:1}
For simplicity, we consider a single complex field $\phi$ with chemical potential $\mu$ coupled to the $U(1)$ current associated with the global transformation $\phi \to e^{i \alpha} \phi$.
Generalizing the following results to the case of $O(N)$ with fundamental matter is straight-forward, since the chemical potential always couples to $U(1)$ charges embedded within $O(N)$.
If N is even, then with an appropriate change of basis, the theory can be written in terms of $N/2$ complex scalar fields---each of which is coupled to an independent chemical potential.
The techniques discussed for a single complex scalar field can then be applied to each of the $N/2$ fields individually.
If $N$ is odd, then there will be an additional real, uncharged scalar field which plays no role in the following discussion.

For a single complex scalar field, the grand canonical partition function in Euclidean space-time takes the generic form
\beq
Z(\mu) = \int\! [d\phi^\ast]\,[d\phi]\, e^{-S_0[\phi^\ast,\phi]-S_1[\phi^\ast,\phi]}\ ,
\eeq
where $S_0$ and $S_1$ correspond to the free and interacting parts of the action respectively.
The specific form of $S_0$ will depend on whether one is interested in relativistic or nonrelativistic scalar theories;
both cases will be discussed in detail in the following subsections.
We focus primarily on the free part of the action at nonzero $\mu$, though interactions and additional flavors may be added in a trivial manner.
We assume that $S_1$ is symmetric under $U(1)$ {\it gauge} transformations for reasons which will become clear later on.

We regulate the $(d+1)$-dimensional continuum theory by applying the usual lattice discretization procedure.
We set the spatial and temporal lattice constants equal and work on a $\Ns^d \times \Nt$ lattice with an inverse temperature given by $\beta = \Nt$.
For notational convenience, all dimensionful parameters are written in units of the lattice constant.
Continuum derivatives are replaced with finite differences $\partial_\nu \phi \to \phi_\bfn-\phi_{\bfn-\bfe_\nu}$, and following the prescription of \cite{Hasenfratz:1983ba}, we introduce the chemical potential as the imaginary zero component of the vector potential (i.e. $A_\nu \to i \mu \delta_{\nu,0}$).
This prescription not only avoids spurious cutoff dependence in thermodynamic observables, but is also a natural choice within our formalism, enabling us to draw a direct connection with the fugacity expansion
\beq
Z(\mu) = \sum_{Q\in{\mathbb Z}} Z_Q e^{\beta\mu Q}\ .
\eeq
Throughout the paper we impose periodic boundary conditions in the space and time directions.

We begin by applying the dual lattice transformation to the partition function for free relativistic and nonrelativistic scalars.
Later, we introduce interactions as well as repeat the derivation for expectation values of observables.
A general discussion on character expansions can be found in \cite{Ukawa:1979yv,Creutz:1984mg}.

\subsection{Relativistic theory}
\label{sec:1a}
The free relativistic Euclidean lattice action is defined by
\beq
S_R = \sum_\bfn \left[\sum_\nu \left(2 \phi_\bfn^\ast \phi_\bfn-\phi_\bfn^\ast e^{-\mu\delta_{\nu,0}} \phi_{\bfn+\bfe_\nu}-\phi_{\bfn+\bfe_\nu}^\ast e^{\mu\delta_{\nu,0}} \phi_\bfn \right) + m^2 \phi_\bfn^\ast \phi_\bfn \right]\ ,
\eeq
where $\nu$ runs over positively oriented basis vectors.
In terms of the polar coordinates $\phi = \rho e^{i \theta}$, the partition function may be expressed as
\beq
Z_R(\mu) = \int_0^{2 \pi}\! [d\theta]\, \int_0^\infty\![\rho d\rho]\, \prod_\bfn \left[e^{-\left(2d+2+m^2\right)\rho_\bfn^2}
           \prod_\nu e^{2 \rho_\bfn \rho_{\bfn+\bfe_\nu} \cos(\theta_\bfn-\theta_{\bfn+\bfe_\nu}-i\mu\delta_{\nu,0})}  \right]\ .
\eeq
As previously stated, performing the $\theta$-integration analytically will yield a new representations for the partition function,  expressed as a sum over integer valued link variables as well as the radial degrees of freedom.
To facilitate the integration over angular degrees of freedom, we make use of the convergent series expansion
\beq
\label{eq:BesselSum}
e^{x \cos(z)} = \sum_{l\in \mathZ} I_l(x) e^{i l z}\ ,
\eeq
where $I_l(x)$ is the modified Bessel function of order $l$, with $x\in{\mathbb R}$ and $z\in{\mathbb C}$.
Inserting \Eq{BesselSum} into the partition function and exchanging the sum over modified Bessel functions with the product over nearest neighbors, we obtain
\beq
\label{eq:UglyMess}
Z_R(\mu) = \int_0^{2 \pi}\! [d\theta]\, \int_0^\infty\![\rho d\rho]\, \sum_{\{l_{\bfn,\nu}\}\in\mathZ}
           \prod_{\bfn} \left[e^{-\left(2d+2+m^2\right)\rho_\bfn^2}\prod_\nu I_{l_{\bfn,\nu}}(2 \rho_\bfn\rho_{\bfn+\bfe_\nu}) 
           e^{i l_{\bfn,\nu}(\theta_\bfn-\theta_{\bfn+\bfe_\nu}-i\mu\delta_{\nu,0})}\right]\ ,
\eeq
where the sum over integer valued link variables $l$ correspond to the sum over characters of $U(1)$ (or Fourier modes).
For the moment, the sum over link variables in equation \Eq{UglyMess} is unconstrained.
However, upon integration over $\theta$ one finds that the only nonvanishing contributions to the partition function are those terms {\it independent} of $\theta$.
These contributions may be characterized by link field configurations that satisfy the continuity equation
\beq
\label{eq:continuity}
\partial \cdot l_\bfn \equiv \sum_\nu \left(l_{\bfn,\nu}+l_{\bfn,-\nu}\right) = 0\ ,
\eeq
where $l_{\bfn,\nu} \equiv -l_{\bfn+\bfe_\nu,-\nu}$.
The partition function takes the final form
\beq
\label{eq:RelativisticDualZ}
Z_{R}(\mu) = \sum_{\substack{\{l_{\bfn,\nu}\}\in\mathZ \\\partial\cdot l_\bfn = 0 }} \int_{0}^{\infty}\! [\rho d\rho]\, e^{-\tilde S_R[\rho,l]+\mu \sum_\bfn l_{\bfn,0}}\ ,
\eeq
where
\beq
\label{eq:RelativisticDualAction}
e^{-\tilde S_R[\rho,l]} = \prod_\bfn \left[e^{-(2d+2+m^2)\rho_\bfn^2}\prod_\nu I_{l_{\bfn,\nu}}(2 \rho_\bfn\rho_{\bfn+\bfe_\nu})  \right]\ .
\eeq

The constraint given by \Eq{continuity} allows us to identify the link field $l$ as a conserved current associated with the $U(1)$ symmetry of this theory.
This observation will be justified later when we introduce gauge interactions.
Furthermore, the chemical potential now couples to the the total charge $Q_0 = \beta^{-1}\sum_\bfn l_{\bfn,0}$.

A key property of the modified Bessel functions appearing in \Eq{RelativisticDualAction} is that $I_l(x) \geq 0$ for all $x \geq 0$ and $l\in{\mathbb Z}$.
The partition function is therefore a sum over real and positive weights, rendering the theory suitable for Monte Carlo simulations.
We would like to emphasize at this stage that our derivation remains unchanged in the presence of $U(1)$ gauge symmetric interactions because such interactions are necessarily independent of the angular mode $\theta$.

\subsection{Nonrelativistic theory}
\label{sec:1b}
In the nonrelativistic limit, one may proceed in an analogous fashion to the relativistic case.
\footnote{A similar approach was taken by \cite{Hebert:2000cf} for the Bose Hubbard model, starting from the operator formalism. Numerical simulations were performed in the so-called ``hard core'' approximation.}
The free  Euclidean lattice action is defined by
\beq
S_{NR} = \sum_\bfn \left[\phi_\bfn^\ast (\phi_\bfn- e^{\mu} \phi_{\bfn-\bfe_0})+\frac{1}{2m}\sum_j |\phi_\bfn-\phi_{\bfn-\bfe_j}|^2 \right]\ ,
\eeq
where the sum over basis vectors $\bfe_j$ run over the spatial directions.
After performing shifts in the dummy variable $\bfn$ and changing to polar coordinates, the action takes the form
\beq
S_{NR} = \sum_\bfn \left[-\rho_\bfn \rho_{\bfn+\bfe_0} e^{i (\theta_\bfn-\theta_{\bfn+\bfe_0}-i\mu)}
         -\frac{1}{m}\sum_j \rho_\bfn \rho_{\bfn+\bfe_j}\cos(\theta_\bfn-\theta_{\bfn+\bfe_j}) + \left(1+\frac{d}{m}\right) \rho_\bfn^2 \right]\ .
\eeq
Using \Eq{BesselSum} and the relation
\beq
e^{x e^z} = \sum_{l\in\mathN} \frac{x^l}{l!} e^{l z}\ ,
\eeq
we integrate out the angular degrees of freedom and obtain the dual partition function
\beq
Z_{NR}(\mu) = \sum_{\{l_{\bfn,0}\}\in\mathN}\sum_{\substack{\{l_{\bfn,j}\}\in\mathZ\\\partial\cdot l_\bfn = 0 }} \int_{0}^{\infty}\! [\rho d\rho]\, e^{-\tilde S_{NR}[\rho,l]+\mu \sum_\bfn l_{\bfn,0}}\ ,
\label{eq:NonrelativisticDualZ}
\eeq
with the nonrelativistic dual action defined by
\beq
e^{-\tilde S_{NR}[\rho,l]} = \prod_\bfn \left[e^{-(1+d/m)\rho_\bfn^2}\frac{(\rho_\bfn \rho_{\bfn+\bfe_0})^{l_{\bfn,0}}}{l_{\bfn,0}!} 
                                       \prod_j I_{l_{\bfn,j}}\left(\frac{\rho_\bfn \rho_{\bfn+\bfe_j}}{m}\right) \right]\ .
\eeq
As with the relativistic theory, the nonrelativistic current density is constrained by the continuity equation \Eq{continuity}.
In addition, the zero-component of the current density is restricted to non-negative integers, reflecting the fact that nonrelativistic particles may only propagate forward in time.
Once more, the partition function is free of sign problems and is therefore suitable for Monte Carlo studies.

\subsection{Observables}
\label{sec:1c}
We next address the computation of expectation values for both the relativistic and nonrelativistic theories.
We begin by considering the  generic $2k$-point correlation function 
\beq
G_{\bfm_1, \ldots, \bfm_k,\bfn_1, \ldots \bfn_k}^{(2k)} &=& \la \phi_{\bfm_1}^\ast \ldots \phi_{\bfm_k}^\ast \phi_{\bfn_1} \ldots \phi_{\bfn_k} \ra \cr
 &=& \la \rho_{\bfm_1} \ldots \rho_{\bfm_k} \rho_{\bfn_1} \ldots \rho_{\bfn_k} e^{i(\theta_{\bfn_1}+\ldots+\theta_{\bfn_k}-\theta_{\bfm_1}-\ldots-\theta_{\bfm_k}) }\ra\ .
\eeq
Following the formalism in the previous subsections, we integrate over angular variables and obtain
\beq
G_{\bfm_1, \ldots, \bfm_k,\bfn_1, \ldots \bfn_k}^{(2k)} = \frac{1}{Z(\mu)} \sum_{\substack{\{l_{\bfn,\nu}\}\in\mathZ\\\partial\cdot l_\bfn = J_\bfn}} \int_{0}^{\infty}\! [\rho d\rho]\, \rho_{\bfm_1}
     \ldots \rho_{\bfm_k} \rho_{\bfn_1} \ldots \rho_{\bfn_k} e^{-\tilde S[\rho,l]} e^{\mu \sum_\bfn l_{\bfn,0}}\ ,
\eeq
where $\tilde S$ may represent either the relativistic or nonrelativistic action, including the interaction term $S_1$.
The sum over current densities is, however,  no longer divergence free.
In heuristic terms, an insertion of $\phi_\bfn$ creates one unit of charge at site $\bfn$, whereas an insertion of $\phi_\bfm^\ast$ annihilates a unit of charge at site $\bfm$.
We therefore expect the nontrivial divergence condition
\beq
\partial \cdot l_\bfn = \delta_{\bfn,\bfn_1}+\ldots+\delta_{\bfn,\bfn_k}-\delta_{\bfn,\bfm_1}-\ldots-\delta_{\bfn,\bfm_k} \equiv J_\bfn\ ,
\eeq
where $J$ represents a linear combination of sources and sinks.
In general terms, all gauge variant operators will give rise to nontrivial divergence properties of the current density.
We discuss solutions to these constraints in the following section.

Applying the formalism above, it is possible to calculate all thermodynamic observables of interest.
The dual representation is particularly well-suited for calculating current densities, charge distributions and current density correlation functions, since the current density is one of the dynamical degrees of freedom in this representation.
In our numerical studies of relativistic bosons, we pay particular attention to the charge density $\la n \ra$ and susceptibilities $\chi_{ab}$, defined by
\beq
\chi_{ab}=\Nt \Ns^d \left(\la A B \ra - \la A \ra \la B \ra \right)\ ,
\eeq
where $a$ and $b$ represent source terms for the generic operators $A$ and $B$.
For this study, we consider the susceptibilities $\chi_{\mu\mu}$, $\chi_{\mu m^2}$ and $\chi_{m^2 m^2}$ where the source terms $\mu$ and $m^2$ correspond to the operators
\beq
n = \frac{1}{\Nt \Ns^d}\sum_{\bfn} l_{\bfn,0}\ ,
\eeq
and
\beq
\overline{|\phi|^2} = \frac{1}{\Nt \Ns^d} \sum_\bfn \rho_\bfn^2\ ,
\eeq
respectively.

\subsection{Constraints}
\label{sec:1d}

Standard procedures exists for solving the divergence constraints on current densities \cite{Ukawa:1979yv}.
We begin by decomposing the link variables in terms of homogeneous and particular solutions to $\partial \cdot l_\bfn = J_\bfn$.
The homogeneous solutions may me labeled by $l^Q$ and are characterized by $d+1$ integer valued topological charges defined by
\beq
Q_\nu = \sum_\bfn \delta_{\bfn_\nu,0} l^Q_{\bfn,\nu}\ .
\eeq
As previously discussed, $Q_0$ corresponds to the total conserved charge of the system, whereas $Q_j$ corresponds to the current flux in spatial directions.
With this decomposition, expectation values take the generic form
\beq
\la \calO(\phi) \ra = \frac{1}{Z} \sum_{Q\in\mathZ} \sum_{\substack{\left\{l_{\bfn,\nu}^Q\right\}\in\mathZ \\ \partial\cdot l^Q_\bfn = 0}} \int_{0}^{\infty}\! [\rho d\rho]\, \calO(\rho)
       e^{-\tilde S[\rho,l^Q+l^J]} e^{\mu\sum_\bfn (l_{\bfn,0}^Q+l_{\bfn,0}^J)}\ .
\label{eq:genexp}
\eeq

\begin{figure}
\centering
\includegraphics[width=\figwidth]{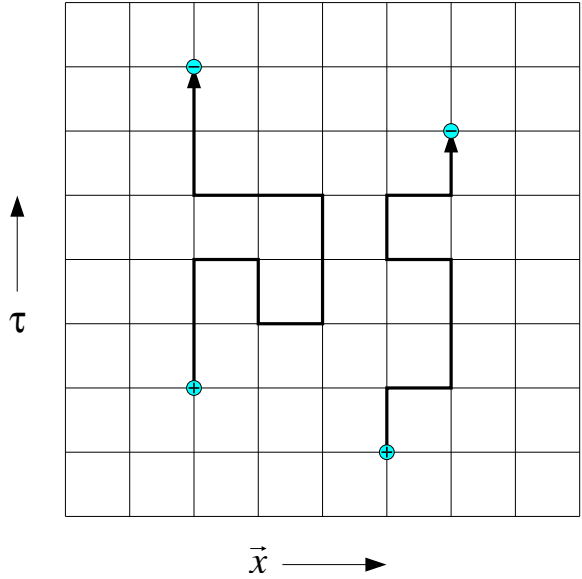}
\caption{
A particular solution to $\partial \cdot l_\bfn = J_\bfn$ arising from the four-point correlation function $G^{(4)}$.
Bold link variables equal $+1$($-1$) if the link orientation and basis vector are parallel (antiparallel).
All remaining links equal zero.
}
\label{fig:strings}
\end{figure}

Note that the particular solution $l^J$ is not unique; one solution may be related to another by a suitable change of variables.
A convenient choice for $l^J$ is to set all links equal to zero except for a set of ``strings'' which connect source-sink pairs as shown in \Fig{strings}.
In the relativistic theory, such strings may flow either forward or backward in time, and may wrap completely around the space-time torus; this would correspond to an increase or decrease in the total charge of the system by one unit (note that this may be undesirable in the canonical ensemble).
In the case of nonrelativistic theories, we shall assume for simplicity that such strings only flow forward in time and do not wrap entirely around the space-time torus.

\begin{figure}
\centering
\includegraphics[width=\figwidth]{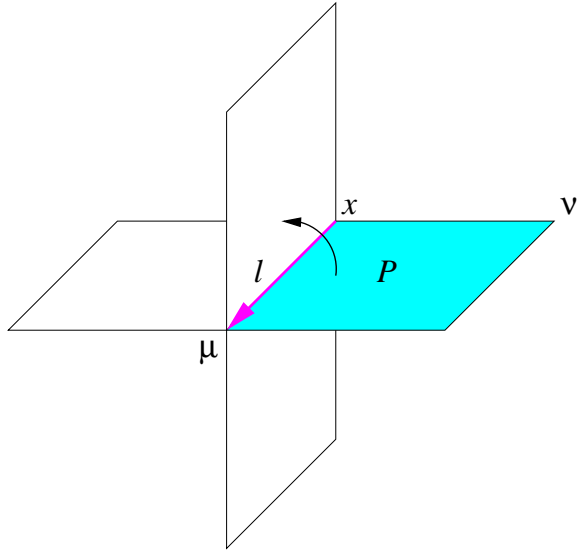}
\caption{
Solving zero divergence constraints by introducing integer valued plaquette variables.
The sum over all (oriented) plaquettes $P$ circulating a common link defines the value of link variable $l$.
}
\label{fig:plaquettes}
\end{figure}

The zero divergence constraint on on link field configurations in the $Q=0$ sector may be solved by changing to plaquette variables $P_{\bfn,\mu\nu} = -P_{\bfn,\nu\mu} \in \mathZ$.
A plaquette is specified by a site $\bfn$ and the two (different) basis vectors $\mu$ and $\nu$ that span it and is positively oriented if $\mu < \nu$.
The divergence free link variables in the $Q=0$ sector are given by
\beq
\label{eq:LatticeCurl}
l_{\bfn,\mu}^{Q=0} = \sum_{\nu \neq \mu} \left(P_{\bfn,\mu\nu} + P_{\bfn,\mu-\nu}\right)\ ,
\eeq
which is depicted graphically in \Fig{plaquettes}.
Note that \Eq{LatticeCurl} has a continuum analog; in three dimensions, for instance, a divergence-free vector field may always be written as the curl of an unconstrained vector field.
Similar relations hold in higher dimensions as well, in the language of differential forms.
Configurations associated with the $Q \neq 0$ sectors may be obtained from the $Q=0$ configuration by increasing the value of all link fields belonging to a closed loop around the space-time torus by $Q_\nu$ units in the $\nu$ direction.

\subsection{$U(1)$ gauge fields}
\label{GaugeFields}
In the presence of gauge interactions, the sign problem reemerges within our formalism.
In the compact $U(1)$ gauge theory (scalar electrodynamics), the vector potential $A$ couples to the integer valued conserved current $l$.
This result is easily verified by repeating the derivation in the preceding subsections, which yields the gauge invariant interaction
\beq
\tilde{S} \supset i \sum_\bfn\sum_\nu l_{\bfn,\nu} A_{\bfn,\nu}\ .
\eeq
Note that the form of the gauge coupling confirms our interpretation of the link field as a current density.
As one may expect, by going to the dual formulation we have traded the chemical potential sign problem for a vector potential sign problem.
A natural remedy is to apply the character expansion to the gauge partition function in addition to the scalar partition function.
Such transformations are not new in the context of gauge theories; character expansions have played an important role in understanding the strong coupling behavior of such theories.
In the case of lattice Yang-Mills theories with Wilson-type actions, the character expansion has a finite radius of convergence \cite{Osterwalder:1977pc}, which likely limits the utility of this discussion.

For completeness, we briefly describe the application of character expansions to dynamical $U(1)$ gauge fields coupled to the conserved current $l$.
Consider the compact $U(1)$ gauge theory defined by the Wilson action
\beq
Z_{gauge} = \int_0^{2\pi}\! [d A]\, e^{\frac{1}{g^2}\sum_\bfn\sum_{\mu\nu}\cos\Theta_{\bfn,\mu\nu} +i \sum_\bfn\sum_\nu l_{\bfn,\nu} A_{\bfn,\nu}}\ ,
\eeq
where $\Theta_{\bfn,\mu\nu} = A_{\bfn,\mu}+A_{\bfn+\bfe_\mu,\nu} - A_{\bfn+\bfe_\nu,\mu}-A_{\bfn,\nu}$ is the lattice field strength tensor.
The gauge partition function vanishes unless the topological charge $Q_\nu$ associated with the conserved current $l$ equals zero.
After applying the character expansion to the gauge partition function, we obtain
\beq
\label{eq:ZGauge}
Z_{gauge} = \sum_{\substack{\{F_{\bfn,\mu\nu}\}\in\mathZ\\\partial\cdot F_\bfn = l_\bfn}} \prod_{\bfn}\prod_{\mu\nu} I_{F_{\bfn,\mu\nu}}\left( \frac{1}{g^2} \right)\ ,
\eeq
where the integer valued plaquette variables $F$ satisfy
\beq
\label{eq:GaugeContinuity}
(\partial\cdot F_\bfn)_\mu = \sum_{\nu}\left(F_{\bfn,\mu\nu}+F_{\bfn,\mu-\nu}\right) = l_{\bfn,\mu}\ .
\eeq
This constraint is simply the lattice analog of the continuum equation of motion for the gauge field.
Note that the partition function \Eq{ZGauge} implicitly depends on the current density $l$ though the \Eq{GaugeContinuity}.

In the absence of current sources, there are $d (d-1)/2$ topological charges associated with the plaquette variables defined by
\beq
Q_{\mu\nu} = \sum_\bfn \delta_{\bfn_\nu,0} \delta_{\bfn_\mu,0} F_{\bfn,\mu\nu}\ .
\eeq
These charges may be visualized as two-dimensional oriented ``sheets'' that wrap around the space-time torus.
The source-free constraint on the plaquette field $F$ may then be solved by a change of variables to integer valued cubes.
Inhomogeneous solutions to $\partial\cdot F_\bfn = l_\bfn$ may be obtained by joining all current loop source-sink pairs with flux tubes, in an analogous manner to our treatment of current sources and sinks.

\section{Algorithms}
\label{sec:2}
\subsection{Loop algorithm}
The dual partition function for scalar theories involves an integral over site variables $\rho$ and a constrained sum over the link field $l$.
As discussed in the previous section, the constraint on $l$ is solved by making an appropriate change of variables.
For the purpose of numerical simulation one need not work directly with plaquette variables, however.
An easier and equivalent approach is to simply consider fluctuations in conserved current loops.
\begin{figure}
\centering
\includegraphics[width=\figwidth]{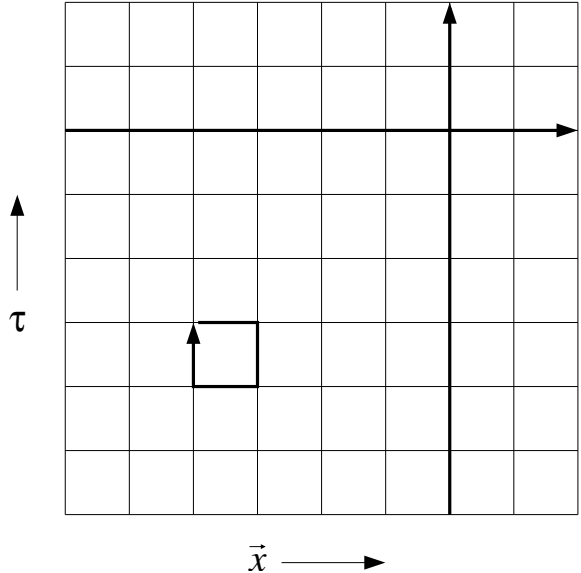}
\caption{
Local (lower left) and global (upper right) current loops on a periodic lattice.
Bold links are updated by +1(-1) if the link orientation and basis vector are parallel (antiparallel).
}
\label{fig:currentloops}
\end{figure}
We consider two types of current loop updates: local loops shown in the lower left of \Fig{currentloops}, and global current loops shown in the upper right of \Fig{currentloops} which correspond to an increase or decrease in the charge of the system by one unit.
Note that an algorithm based on local and global loop updating is ergodic: any divergence-free configuration with charge $Q$ may be obtained from any other with like charge by a sequence of local current loop updates.
Divergence-free configurations in different topological sectors may be related by global current loop updates, possibly succeeded by a sequence of local loop updates.

One may use the standard Metropolis \cite{Metropolis:1953am} accept/reject method for local site and loop, as well as global loop updates.
However, as lattice volumes become large, the acceptance rate for the global loop update is expected to drop to unacceptably low levels.
This in turn adversely affects the correlation time for configurations within the ensemble.
The low acceptance rate is due to the fact that the charge transition probability scales as $e^{-L}$, where $L\sim\Nt,\Ns$ is the linear length of the global current loop.
The likelihood for the system to become stuck in charge sectors which are energetically disfavored increases with volume sizes; this issue may be particularly problematic during the equilibration stage of a Monte Carlo simulation.

The problem of slow tunneling rates may be handled in several different ways.
In the the thermodynamic limit, fluctuations in the charges $Q_j$ may be neglected altogether since these fluctuations will have little influence on local observables.
$Q_0$ fluctuations may be neglected for similar reasons at zero temperature, as well as in cases where canonical ensemble simulations are of interest.
At finite but large temperatures and in the grand canonical ensemble, one might avoid the problem by implementing an intermediate equilibration step to facilitate fluctuations between different $Q_0$ sectors, followed by an accept/reject step.

\subsection{Worm algorithm}

An alternate method for avoiding long correlation times associated with charge fluctuations is the ``worm'' algorithm \cite{Prokofev:2001gh}.
This algorithm may be directly applied to the partition functions given by  \Eq{RelativisticDualZ} and \Eq{NonrelativisticDualZ} which were derived from the character expansion in addition to partition functions derived from hopping parameter expansions.
We will focus on the former; the later is discussed in \cite{Prokofev:2001gh} at zero chemical potential and may be generalized to dense matter in a straight-forward manner.
We emphasis that in each of the above cases the role played by the chemical potential is the same.

The worm algorithm employs a dynamical source and sink associated with the current density to generate link field configurations.
The process begins by associating a current source and sink with a randomly chosen site on the lattice.
One then attempts to move the source to a neighboring site by randomly updating one of the $2(d+1)$ link fields (using Metropolis accept/reject method) associated with the bond by one unit.
As links are updated, the divergence condition on the link field becomes $\partial\cdot l_\bfn = \delta_{\bfn,\bfn_a} -\delta_{\bfn,\bfn_b}$, where $\bfn_a$ and $\bfn_b$ represent the respective locations of the source and sink.
Proceeding one link at a time, the source traces out a string or worm on the space-time lattice.
Most of the link field configurations generated by this method are unphysical and may be discarded because they do not satisfy the divergence-free constraint.
If the endpoints of the string occupy the same site, then the divergence-free constraint is satisfied and the field configuration is stored for later calculation of physical observables.
A new site is then chosen at random and the process is repeated.

At finite density, the probability for the source to propagate forward in imaginary time is enhanced by a factor of $e^\mu$.
This bias magnifies the likelihood for the source to traverse the temporal extent of the lattice after a sequence of updates, resulting in an increase or decrease of the topological charge $Q_0$ by one unit. 
The main advantage of this algorithm is that tunneling processes are avoided.
Instead of going over the energy barrier associated with changes in $Q_0$, the worm algorithm allows one to circumnavigate the barrier by accessing unphysical states of the system--states where the charge $Q_0$ is not well-defined.

\subsection{The gauge sector}

\begin{figure}
\centering
\includegraphics[width=\figwidth]{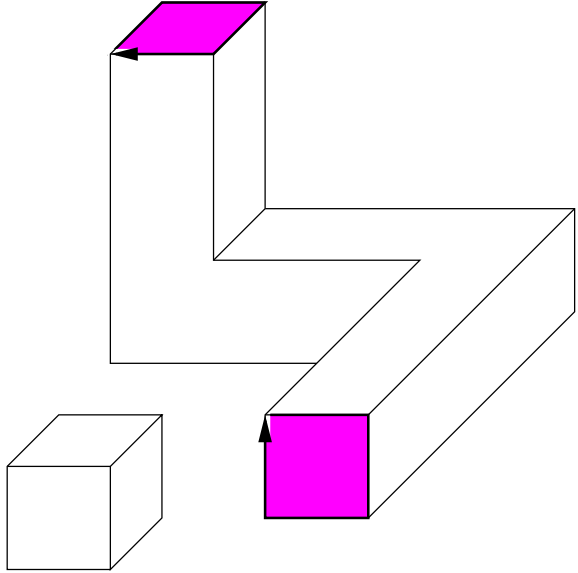}
\caption{
Local flux tube (left) and flux tube connecting a current loop source and sink (upper right).
}
\label{fig:fluxtubes}
\end{figure}

In numerical simulations involving both gauge and matter fields, one must allow fluctuations in link variables $l$ and plaquette variables $F$, subject to constraints.
The issues may be dealt with in an analogous fashion to the loop algorithm.
The divergence constraint on the plaquette field is satisfied by considering fluctuations in local flux tubes (cubes).
On small lattices, fluctuations in the global charge $Q_{\mu\nu}$ must also be considered.
Since these charges correspond to sheets (with area $A$), the probability for fluctuations of this kind scale as $e^{-A}$.
In addition to these, one must consider fluctuations involving gauge plaquettes bounded by current loops; these current loops act as flux tube sources and sinks.
An example configuration is displayed in \Fig{fluxtubes}.

\subsection{Reweighting}

Before efficient simulations can be performed, additional issues must be addressed.
Ideally one would like to generate a single set of configurations $\{\Phi_i\} = \{\rho_i,l_i \}$, where $i=1,\ldots,\calN$, and $\calN$ is the sample size, distributed according to the probability weight $e^{-S[\Phi]}$.
From this ensemble, one would then like to approximate the expectation values of {\it all} observables using the relation
\beq
\la \calO(\Phi) \ra = \frac{1}{\calN} \sum_{i=1}^{\calN} \calO(\Phi_i) + O\left(\frac{1}{\calN} \right)^{1/2}\ .
\eeq
Unfortunately, in the dual representation link variables must satisfy specific divergence constraints which are governed by the {\it gauge} transformation properties of each observable.
As a result, configurations generated in a simulation may only be used to calculate expectation values of those observables which transform similarly under gauge transformations.
With the use of reweighting \cite{Ferrenberg:1988yz}, this issue may be avoided.
Let $\la \calO(\rho) \ra_J$ represent the expectation value of some $\rho$-dependent operator in the presence of source $J$.
Then referring to \Eq{genexp} one can show that
\beq
\la \calO(\rho) \ra_J &=& \left\langle \frac{ \calO(\rho) e^{-\tilde{S}(\rho,l^Q+l^J)+\mu\sum_\bfn l_{\bfn,0}^J}}{e^{- \tilde{S}(\rho,l^Q)} }  \right\rangle_0 \ ,
\label{eq:reweight}
\eeq
where $\la \ldots \ra_0$ is the expectation values with respect to source-free configurations.
In this scheme, all expectation values may be computed from the same source-free distribution.
One disadvantage of reweighting, however, is that one must now evaluate expectation values of nonlocal observables.
Reweighting is expected to give reliable results provided the Monte Carlo distribution and target distribution have sufficient overlap.
In particular, the extent of correlation functions should remain well within the bulk of the lattice.
Otherwise, the Monte Carlo and true distributions may lie over different topological sectors, leading to large systematic errors.
One may be able to reduce the systematic errors arising from the overlap problem by taking advantage of the fact the the particular solution $l^J$ is not unique.
In particular, one could average the left hand side of \Eq{reweight} over several different particular solutions $l^J$, thereby improving estimates of the expectation value.
While numerical results will most likely improve, the systematic errors remain uncontrolled with this scheme.

\section{Simulations at zero and weak coupling}
\label{sec:3}
We have now provided a framework for studying scalar theories at finite chemical potential on the lattice, free of sign problems.
We test the convergence of our algorithms by performing Monte Carlo simulations of the $(2+1)$-dimensional free, relativistic $U(1)$ theory at finite density and finite temperature.
The algorithm is then applied to the same theory with an additional repulsive interaction
\beq
S_1 = \sum_\bfn \frac{\lambda}{4} |\phi_\bfn|^4\ .
\label{eq:interaction}
\eeq
Here, we determine the $m^2-\mu$ phase diagram at infinite volume and zero temperature, and in a regime where perturbation theory is reliable.

Since these simulations are exploratory in nature, we work in $2+1$ dimensions and limit our lattice sizes to $10^2 \times 10$ or smaller, employing the loop algorithm discussed above.
During each run, we update the system through $\sim 2 \times 10^6$ Monte Carlo steps (1 Monte Carlo step $=$ 1 sweep through the system) and estimate that the correlation times are on the order of $\tau_{corr} \sim 2 \times 10^2$ steps.
Following an equilibration time $\tau_{eq} \sim 2 \times 10^4$, configurations are accepted every $\tau_{corr}$ steps to ensure that ensembles are uncorrelated.
We include fluctuations in global charges $Q_\nu$ to ease the comparison of numerical results with analytic calculations.

\subsection{Free theory}
We begin by measuring properties of the free theory at finite chemical potential.
The free theory is exactly soluble because the action is a quadratic form involving $\phi$.
At zero coupling and finite lattice spacing, the theory is stable provided $4\sinh^2(\mu/2) < m^2$.
At finite volume, the free energy of the system is given by
\beq
e^{-\Nt \Ns f_0(\mu)} &=& \prod_p \frac{1}{m^2 + 4 \sin^2(\frac{p_\nu+i\mu\delta_{\nu,0}}{2})} \cr
                    &=& \prod_{\mathbf{p}} \frac{1}{\cosh(\Nt\cosh^{-1}\xi_{\mathbf{p}})-\cosh(\Nt \mu)}\ ,
\eeq
where
\beq
\label{eq:xi}
\xi_{\mathbf{p}} = 1+\half\left(m^2+4\sum_j \sin^2 \left( \frac{p_j}{2} \right) \right)\ ,
\eeq
and the momenta take the discrete values $p_\nu = 2 \pi n_\nu/N_\nu$ ($N_0 = \Nt$ and $N_j = \Ns$) within the Brillouin zone.
From the free energy we compute the charge density as a function of $\mu$,
\beq
\la n\ra &=& \frac{-\partial f_0(\mu)}{\partial \mu} \cr
  &=& \frac{1}{\Ns^d}\sum_{\mathbf{p}} \frac{\sinh(\Nt \mu)}{\cosh(\Nt\cosh^{-1}\xi_{\mathbf{p}})-\cosh(\Nt \mu)}\ .
\eeq
In the low temperature regime ($\beta m \gg 1$) the charge density remains zero for $4\sinh^2(\mu/2) < m^2$.
At higher temperatures ($\beta m \sim 1$) particles (antiparticles) become thermally excited, leading to finite densities for $|\mu| < m$.
In both cases, as $4\sinh^2(\mu/2) \to m^2$ the charge density $n$ diverges, signaling the onset of Bose-Einstein condensation. 

\begin{figure}
\centering
\includegraphics[width=\plotwidth]{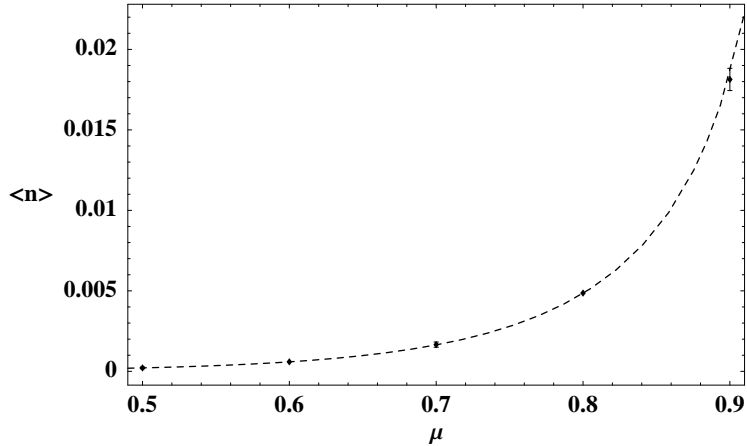}
\caption{
Charge density $\la n\ra$ as a function of $\mu$ for the free relativistic theory ($m=1$) on an $10^2 \times 10$ lattice.
Dashed line indicates the exact analytic calculation, data points indicate numerical results.
}
\label{fig:density}
\end{figure}

\begin{figure}
\centering
\includegraphics[width=\plotwidth]{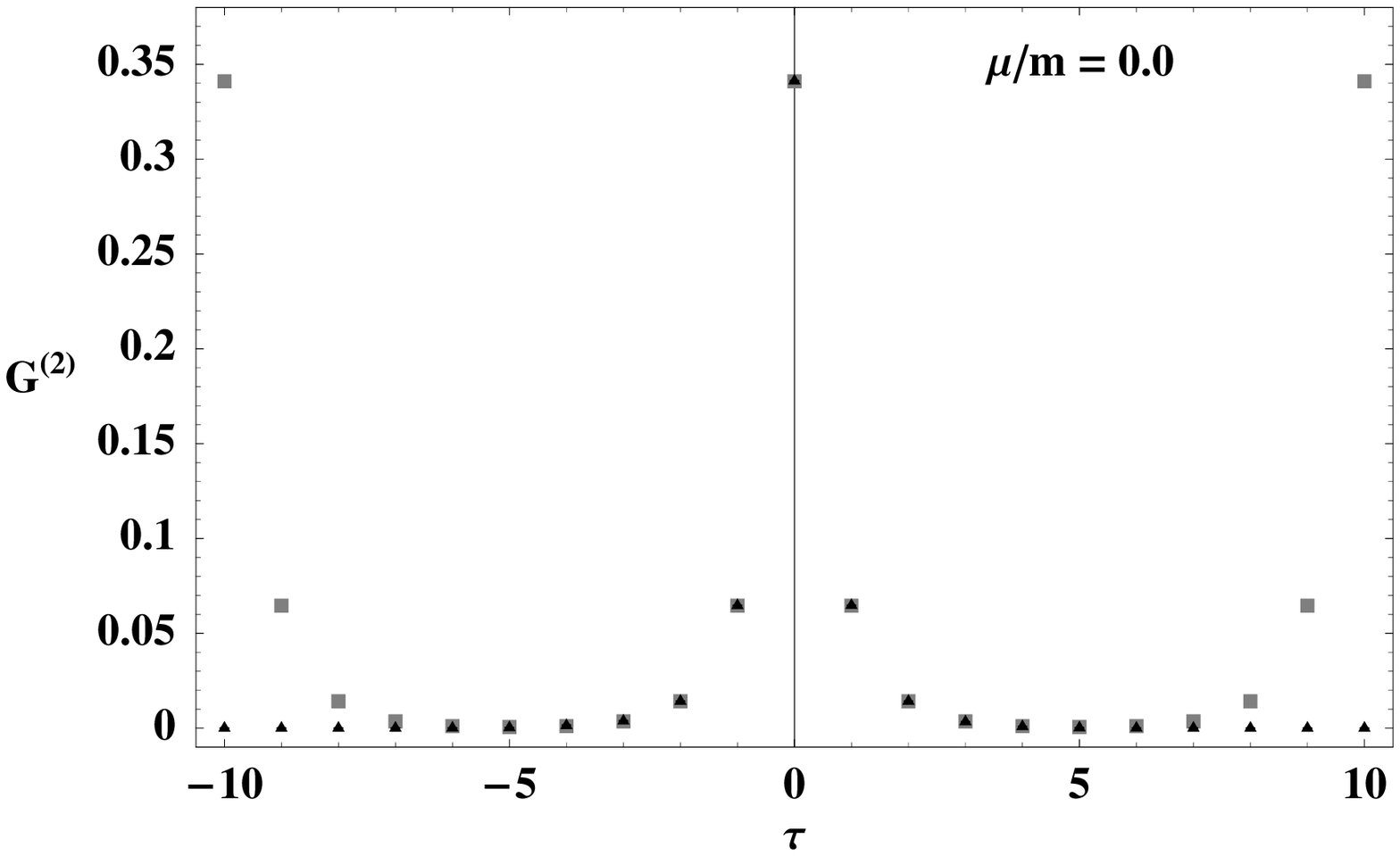} \\
\includegraphics[width=\plotwidth]{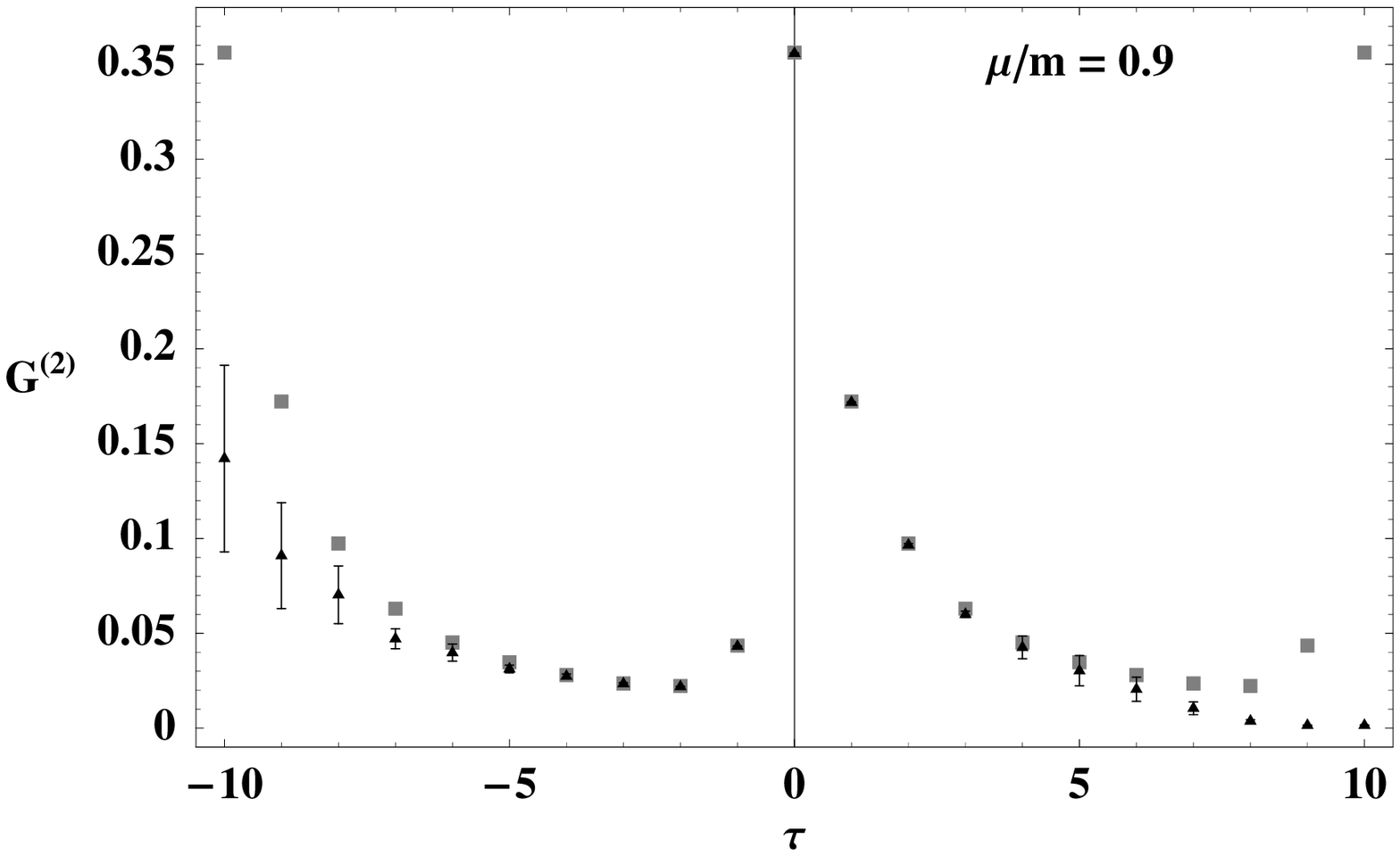}
\caption{
Two point correlation function $G^{(2)}_\tau$ for the free relativistic scalar theory with $\mu/m = 0.0$ and $0.9$.
Error bars indicate only the statistical uncertainty and not that due to systematic errors.
Exact analytic calculations (squares) agree with Monte Carlo simulations (triangles) for $\tau < \Nt/2$.
Results were obtained for a $10^2 \times 10$ lattice
Discrepancies between the numerical data and analytic results for $\tau > \Nt/2$ are discussed in the text.
}
\label{fig:twopt}
\end{figure}

Simulations of the free theory are performed at a fixed  mass ($m = 1$) and for chemical potentials ranging from $\mu = 0.0-0.9$.
\Fig{density} provides a comparison of the exact analytic and numerical computation of the charge density as a function of chemical potential for $\beta m =10$.
In addition, we measure the two-point correlation function $G^{(2)}_\tau = \la \phi^\ast_{\bfn+\tau \bfe_0} \phi_\bfn \ra$ using the reweighting procedure described in \Sec{2}.
Our numerical results for $G_\tau^{(2)}$ in \Fig{twopt} agree with exact calculations for time separations $\tau \ll \Nt$.
For time separations $\tau \gtrsim \Nt$, the reweighting method fails to yield correct results.
This is presumably due to poor overlap between Monte Carlo and target distributions.

\subsection{Interacting theory}
\begin{figure}
\centering
\includegraphics[width=\plotwidth]{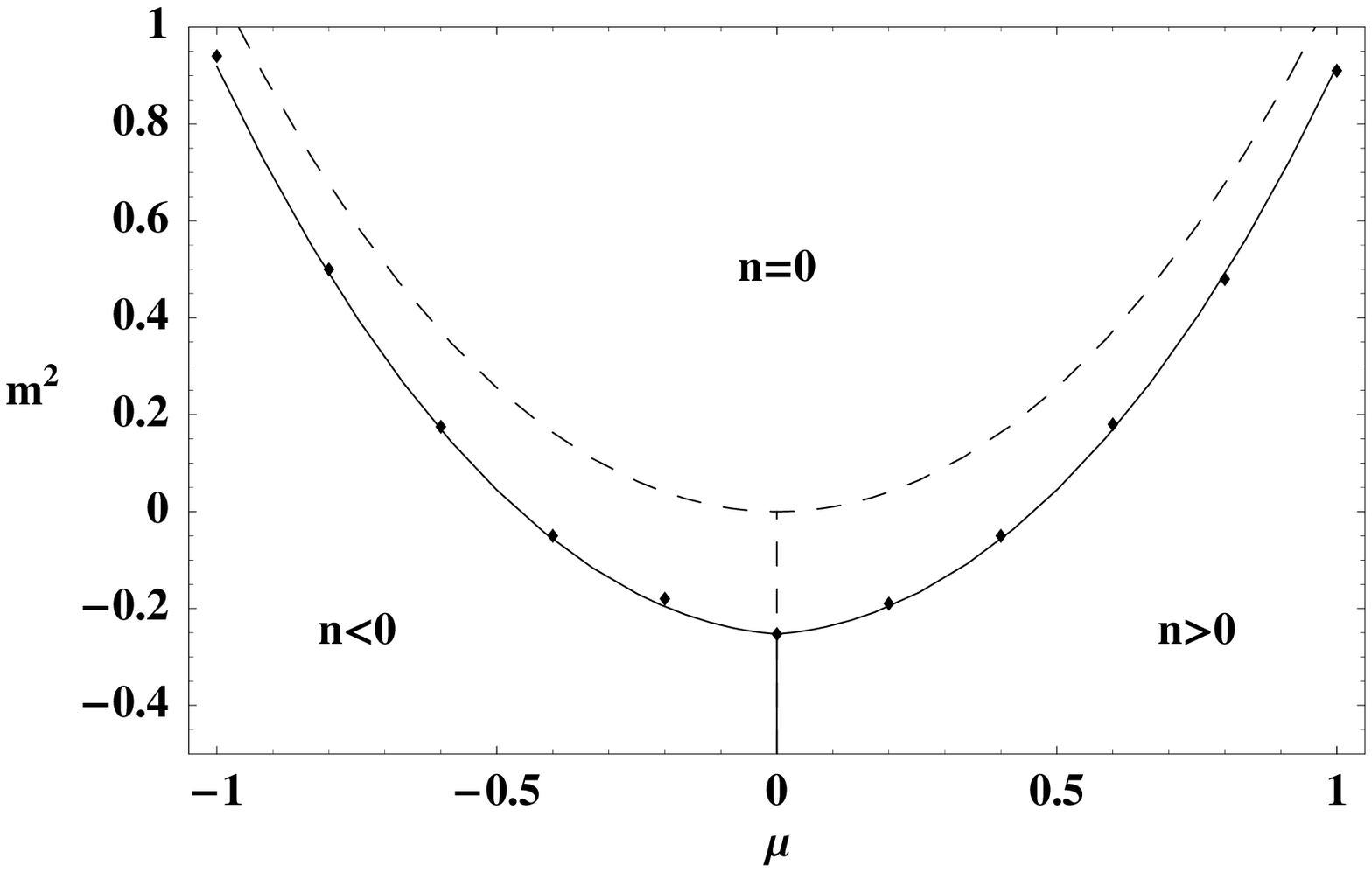}
\caption{
The charge density phase diagram in the $(\mu,m^2)$ plane at $\lambda = 1$.
The dashed (solid) curve corresponds to the tree level (1-loop) calculation of $m_c^2$ as a function of $\mu$.
The upper (lower) region corresponds to the unbroken (broken) phase.
Data points are obtained from the volume dependence of $\chi_{\mu m^2}$, as shown in \Fig{susceptibility}.
}
\label{fig:phase}
\end{figure}

\begin{figure}
\centering
\includegraphics[width=1.5in.]{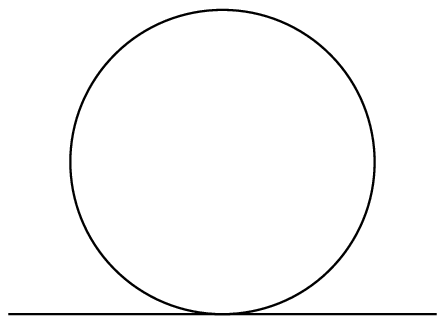}
\caption{
1-loop self energy diagram $\Sigma_1(m^2,\mu)$.
}
\label{fig:selfenergy}
\end{figure}

At weak coupling ($\lambda=1$) and zero temperature we determine the critical mass $m_c^2$ as a function of $\mu$.
The transition occurs when the effective mass  $m_{eff}^2(m^2,\mu) = m^2-4\sinh^2(\mu/2)+\Sigma(m^2,\mu)$ vanishes, where $\Sigma(m^2,\mu)$ represents the selfenergy correction to the free propagator. 
We use this relation to compute $m_c^2$ for fixed values of $\mu$ and $\lambda$.
At tree level, $m_c^2$ is simply given by $m_c^2=4 \sinh^2(\mu/2)$;
this line is indicated by the dashed curve in \Fig{phase}.
We compute the selfenergy perturbatively in the infinite volume limit and at finite lattice spacing.
The first-order correction arising from the 1-loop diagram shown in \Fig{selfenergy} yields the relation $m_c^2 = 4\sinh^2(\mu/2)-\Sigma_1(m_c^2,\mu)$.
The 1-loop  self energy is proportional to the free energy $f_0$ of the noninteracting theory and in the infinite volume limit one finds
\beq
\Sigma_1(m^2,\mu) = \lambda \int_{BZ} \!\frac{d^{d}\mathbf{p}}{(2 \pi)^d}\, \frac{1}{2 \sqrt{\xi_{\mathbf{p}}-1}} 
                  H \left(1-e^{|\mu|}\left(\xi_{\mathbf{p}}-\sqrt{\xi_{\mathbf{p}}^2-1}\right)\right)\ ,
\eeq
where $\xi_{\mathbf{p}}$ is given by \Eq{xi} and $H(x)$ represents the Heaviside step function.
The domain of integration is over spatial momenta $p_j \in [-\pi,\pi]$.
We solve for $m_c^2$ to leading order in $\lambda$ by substituting our tree level result for $m^2$ into $\Sigma_1(m^2,\mu)$.
The first order correction to the critical line is indicated by the solid curve in \Fig{phase}.
Above the critical line ($m>m_c$) the charge density is strictly zero at zero temperature.
Below the critical line ($m<m_c$) the charge density is positive for $\mu>0$ and negative for $\mu<0$.

\begin{figure}
\centering
\includegraphics[width=\plotwidth]{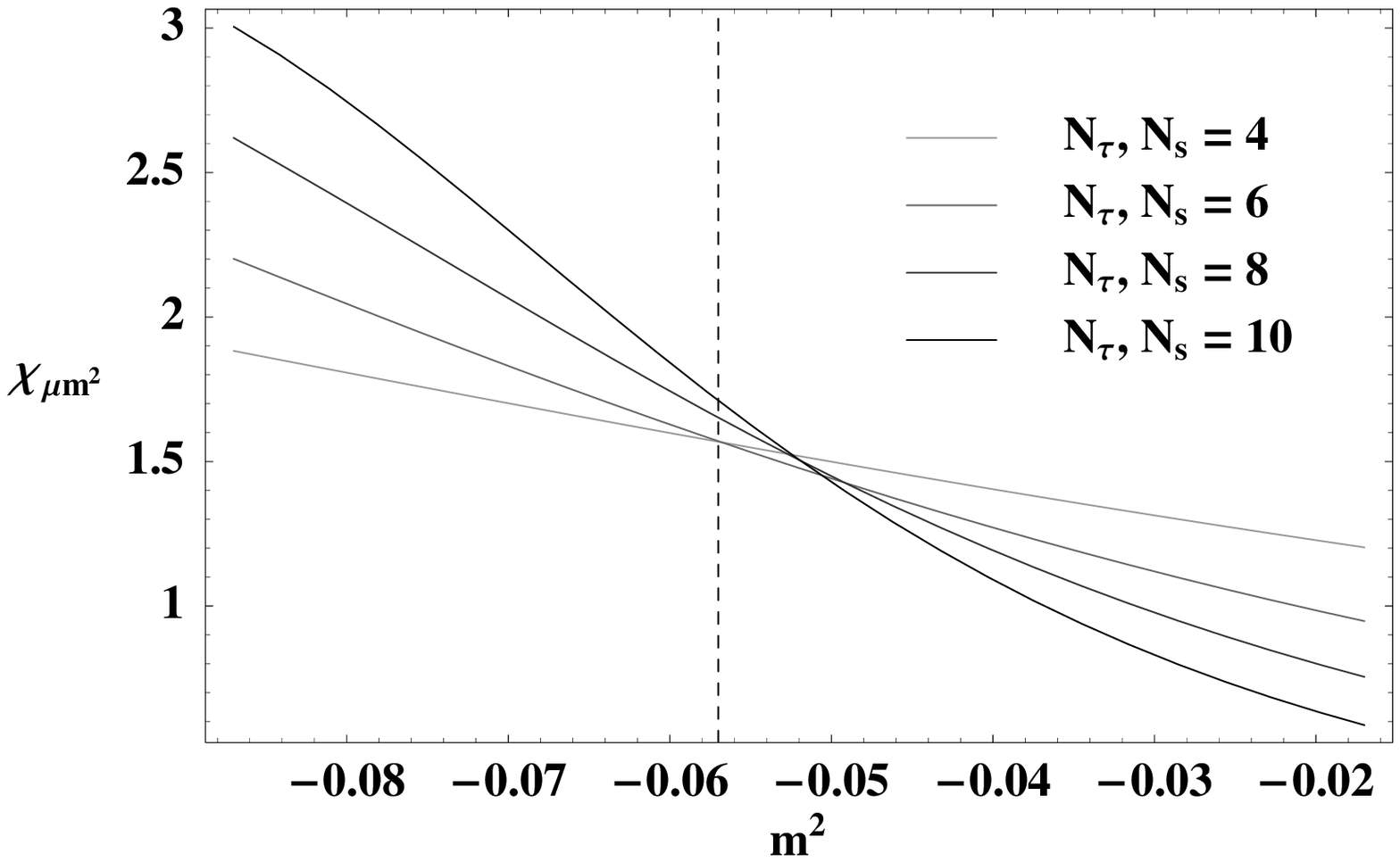}
\caption{
Volume dependence of $\chi_{\mu m^2}$ for $\mu=0.4$ and lattice sizes ranging from $\Nt, \Ns = 4-10$ in the space and time direction.
Curves are obtained by extrapolating data from $m^2=-0.057$ (dashed line) via reweighting.
}
\label{fig:susceptibility}
\end{figure}

In our numerical simulations, the critical mass is determined by observing the dependence of the susceptibilities $\chi_{\mu \mu}$ and $\chi_{\mu m^2}$ on the linear size $\Nt, \Ns$ of our lattice.
In the infinite volume limit, these quantities are discontinuous (assuming $\mu \neq 0$) across the phase boundary.
At finite volume, however, it is well known that no phase transition exists;
discontinuities in thermodynamic quantities become smooth cross-overs.
We obtain an approximation to the infinite volume critical mass by plotting $\chi_{\mu \mu}$ and $\chi_{\mu m^2}$ as functions of $m^2$ for various lattice sizes.
A typical set of data is displayed in \Fig{susceptibility} for $\mu = 0.4$.
Simulations are performed at a trial mass $m_0^2$ near $m_c^2$.
Susceptibilities in the neighborhood of $m_0^2$ are then obtained from the same ensemble with the use of
\beq
\la\calO\ra_{m^2} = \frac{\la\calO e^{-(m^2-m_0^2)\overline{|\phi|^2}}\ra_{m_0^2}}{\la e^{-(m^2-m_0^2)\overline{|\phi|^2}}\ra_{m_0^2}}\ .
\eeq

In the limit $\Nt, \Ns\to\infty$, the susceptibility curves should intersect at the same point, revealing the infinite volume value of the critical mass.
In practice, one should apply a finite scaling analysis to accurately determine $m_c^2$ in the thermodynamic limit.
Since our lattices are relatively small (scaling analysis is only valid for $\Nt, \Ns \gg 1$) and our computing resources limited, we choose not to do so here.
Estimates of $m_c^2$ are obtained from the $\Nt, \Ns = 4, 6, 8,$ and $10$ crossings.
In the special case $\mu =0$, the average number density and susceptibilities remains zero across the phase boundary.
Here, we study the the properties of the charge distribution associated with $Q_0$ as a function of lattice volume.
In particular, in the infinite volume limit the Binder cumulant $U \equiv 1- \la Q_0^4 \ra/3 \la Q_0^2 \ra^2$ equals zero in the broken phase ($m^2<m_c^2$) and diverges in the unbroken phase ($m^2>m_c^2$).
Our numerical results for the $m^2-\mu$ phase diagram shown in \Fig{phase} appear consistent with perturbation theory at weak coupling.

\section{An application}
\label{sec:4}
We now turn to an application, focusing for simplicity on the relativistic $U(1)$ lattice theory with the interaction given by \Eq{interaction} and in a nonperturbative regime.
The purpose of this section is to show that full scale simulations are feasible using the methods outlined above.
Since our path integral representation is derived in the imaginary-time formalism, the utility of our approach is limited to systems in thermodynamic equilibrium.
Nonetheless, the algorithms may, for instance, be used to study the universal properties of the $(2+1)$-dimensional relativistic $|\phi|^4$ theory, which is known to exhibit a second order phase transition at zero temperature and density.
At $\beta, V = \infty$ and at the phase transition, the theory exhibits no intrinsic scale.
As a result, one can make predictions about the dependence of observables on the physical parameters of the theory based on universality and the scaling hypothesis.
As an example, dimensional analysis suggests that $\la n\ra = \calB \mu^2$ \cite{Son:2002uz} and $T_c=\calC \mu$ \cite{Son:2006rt} as one moves away from the zero temperature and density critical line.
The proportionality constants $\calB$ and $\calC$ are believed to be universal and nonperturbative, and therefore ideal for calculation by numerical means.

For the calculation of universal constants, we choose the self coupling $\lambda=192$ and determine the critical mass by studying the volume dependence of the Binder cumulant $U(m^2)$ at zero chemical potential.
Results for $U(m^2)$ were obtained using the worm algorithm in order to facilitate charge fluctuations at larger volume sizes, as discussed in \Sec{2}.
Scans in the mass parameter were performed for lattice sizes ranging from $\Nt, \Ns = 20-36$ sites.
\begin{figure}
\centering
\includegraphics[width=\plotwidth]{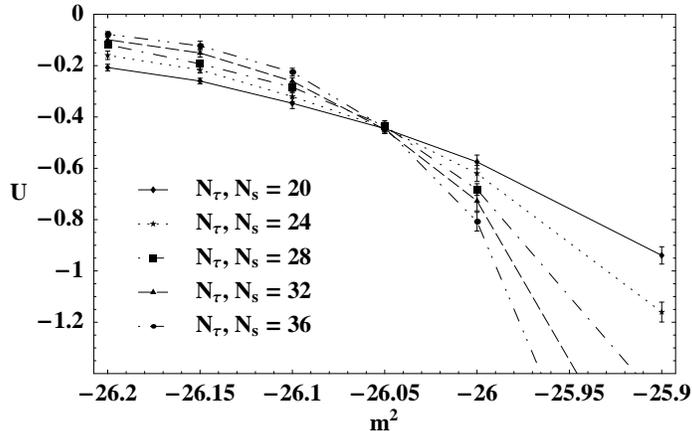}
\caption{
$m^2$ dependence of $U$ for $\lambda=192$,  $\mu=0.0$ and space-time volumes $\Nt, \Ns=20-36$.
}
\label{fig:U}
\end{figure}
The results plotted in \Fig{U} suggest that the critical mass is approximately $m_c^2 \approx -26.05$ for this choice of $\lambda$.

\begin{figure}
\centering
\includegraphics[width=\plotwidth]{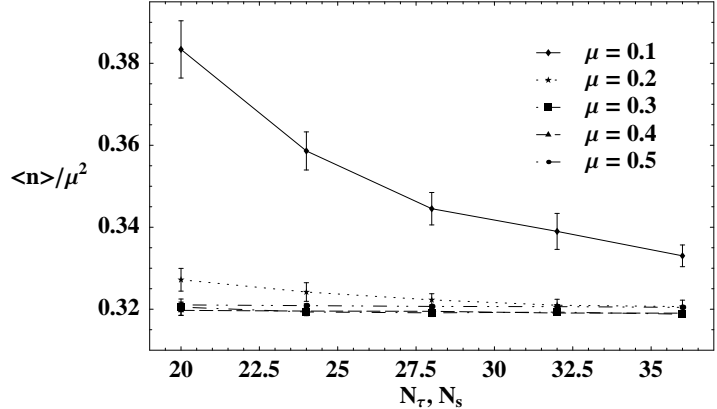}
\caption{
Volume dependence of $\langle n \rangle /\mu^2$ for $\lambda=192$, $m^2 = -26.05$ and $\mu=0.1-0.5$.
}
\label{fig:NvsL}
\end{figure}

The universal constant $\calB$ is determined by studying the volume dependence of the ratio $\la n\ra/\mu^2$ at $\lambda=192$ and $m^2 = m_c^2$.
Curves were obtained by direct measurement of the number density, once again using the worm algorithm.
Scans displayed in \Fig{NvsL} were performed for values of the chemical potential ranging from $\mu=0.1-0.5$ and lattice sizes ranging from $\Nt, \Ns=20-36$ sites.
In the infinite volume limit, $\la n\ra/\mu^2$ is expected to converge to $\calB$.
Previous indirect measurements of this quantity have found $\calB=32(1)$ \cite{Neuhaus:2002fp}.
With exception to the $\mu=0.1$ level curve, we find $\calB\approx0.32$ for large space-time volumes--in apparent agreement with the earlier calculation.

\begin{figure}
\centering
\includegraphics[width=\plotwidth]{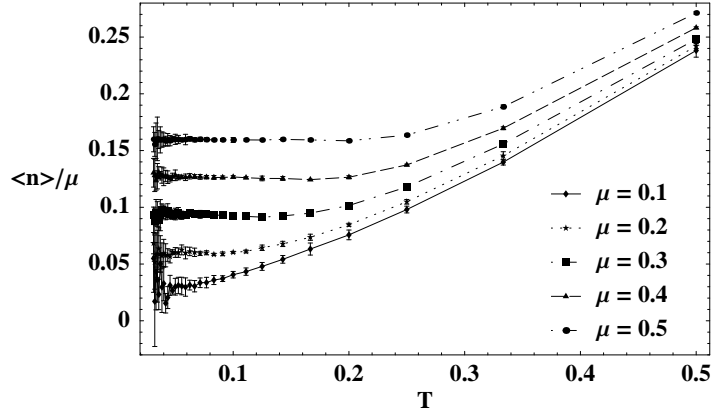}
\caption{
Temperature dependence of $\langle n \rangle/\mu$ for $\lambda=192$, $m^2 = -26.05$, $\mu=0.1-0.5$ and spatial volume $\Ns=16$.
}
\label{fig:NvsT}
\end{figure}

Turning now to the finite temperature behavior near the critical line, we study the temperature dependence of the number density for chemical potentials ranging from $\mu=0.1-0.5$ and using the loop algorithm described in \Sec{2}.
Curves displayed in \Fig{NvsT} were obtained by direct measurement of the number density at a spatial volume $\Ns=16$.
Near the critical point $(\lambda,m_c^2)=(192, -26.05)$, the effective theory describing the Goldstone mode $\theta$ predicts weak temperature dependence in the $T\ll\mu$ regime.
More precisely, using the arguments of \cite{Nishida:2005uf} the leading order contribution to the continuum effective Lagrangian is
\beq
\calL = \frac{\calB}{3} \left[(\partial_\tau \theta + \mu)^2 + (\nabla\theta)^2 \right]^{3/2}\ .
\eeq
Expanding the Lagrangian in inverse powers of $\mu$, we find
\beq
\calL = \frac{\calB}{3}\mu^3 + \calB\mu\left[(\partial_\tau\theta)^2+\half(\nabla\theta)^2 \right]+\ldots\ ,
\eeq
and to leading order in temperature, the number density is given by:
\beq
\la n\ra = \calB\mu^2\left[1+\calO(T/\mu)^4 \right]\ .
\eeq
This predicted temperature dependence is evident in our numerical calculation of $\la n\ra/\mu$ as a function of $T$ for $T\ll\mu$.
Results of this calculation are shown in \Fig{NvsT}.

\begin{figure}
\centering
\includegraphics[width=\plotwidth]{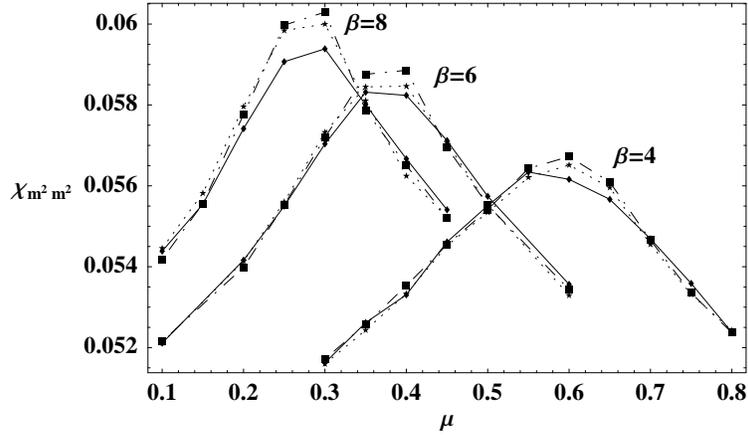}
\caption{
$\mu$ dependence of $\chi_{m^2m^2}$ for $\lambda=192$, and $m^2=-26.05$ for spatial volumes $\Ns=24$ (diamond), $\Ns=36$ (star) and $\Ns=48$ (square).
}
\label{fig:chi}
\end{figure}

For a precise determination of the universal constant $\calC$, which characterizes the shift in $T_c$ as a function of $\mu$, we fix the inverse temperature at $\Nt = 4,6$ and $8$ and study the $\mu$-dependence of the susceptibility $\chi_{m^2m^2}$ for spatial lattice volumes $N_s = 24, 36$ and $48$.
Results are shown in \Fig{chi}.
In the thermodynamic limit $\Ns\to\infty$, this susceptibility is expected to diverge as one approaches criticality.
From the infinite volume location of the maximum $\mu_{\textrm max}(\Ns\to\infty)$ of the susceptibility, one can determine the universal constant by using the relation  $\calC^{-1} = \beta \mu_{\textrm max}(\Ns\to\infty)$.
We leave an infinite volume extrapolation of $\mu_{\textrm max}(\Ns\to\infty)$ to a later study; at finite volume we find the approximate value of $\calC\approx 0.44$ for each choice of $\beta$ and for the spatial volume $\Ns=48$.

\section{Conclusion}
\label{sec:5}
We have derived a new representation for the bosonic partition function which may be interpreted as a path integral over current densities.
The representation avoids the problem of complex phases and is therefore suitable for use in numerical simulations.
The formalism is applicable to relativistic and nonrelativistic $O(N)$ scalar theories at finite density and in the case $N=2$ allows for the inclusion of gauge interactions at strong coupling.
We have verified the method by performing numerical simulations of the free relativistic scalar field theory in $2+1$ dimensions, finding that numerical results for the two-point correlator and charge density agree with exact calculations.
The method was then applied to $|\phi|^4$ theory at weak coupling where we have correctly reproduced the $m^2-\mu$ phase diagram.
Finally, we have demonstrated the utility of the formulation by performing numerical calculations of the universal constants $\calB$ and $\calC$ associated with the phase transition at finite temperature and chemical potential.
We find the approximate values $\calB \approx 0.32$ and $\calC \approx 0.44$; a full analysis, including infinite volume extrapolation of the data is left for future studies.

The methods and ideas discussed in this paper may be extended to include the addition of higher representation fields and multiple flavors.
In cases such as this, positivity of the dual partition function will depend on the specific form of interactions and whether of not the resulting character expansion coefficients satisfy the desired positivity requirements.

Finally, we have been unable to generalize our method to the case of fermionic systems for the obvious reason: no polar decomposition exists for Grassmann variables.
Similar methods which are based on the hopping parameter expansion are known to fail as well because configurations with an an odd number of fermion loops in the expansion carry negative weight.
It is likely that the fermion sign problem is fundamentally different from that for bosons and the techniques describes here are simply inapplicable.

\begin{acknowledgments}
M. G. E. would like to thank  D. B. Kaplan for his guidance throughout the course of this work.
Gratitude is also expressed toward J. E. Drut, S. Sharpe, D. T. Son, M. Wingate and L. Yaffe for their helpful suggestions.
This work was supported by The U.S. Department of Energy Grant DE-FG02-00ER41132.
\end{acknowledgments}

\bibliography{bosons}
\bibliographystyle{apsrev}

\end{document}